\documentclass{aa}
\usepackage{psfig}

\def\bron{SLX~1737$-$282}
\def\ecs{erg~cm$^{-2}$s$^{-1}$}
\def\lum{erg~s$^{-1}$}

\begin{document}

\title{Discovery of the neutron star nature of \bron}

\titlerunning{The nature of \bron} 
\authorrunning{J.J.M. in 't Zand, F. Verbunt, E. Kuulkers et al.}

\author{J.J.M.~in~'t~Zand\inst{1,2}
 \and F.~Verbunt\inst{1}
 \and E.~Kuulkers\inst{1,2}
 \and C.B. Markwardt\inst{3,4}
 \and A.~Bazzano\inst{5}
 \and M.~Cocchi\inst{5} 
 \and R.~Cornelisse\inst{1,2}
 \and J.~Heise\inst{2}
 \and L.~Natalucci\inst{5}
 \and P.~Ubertini\inst{5}
}


\institute{     Astronomical Institute, Utrecht University, P.O. Box 80000,
                NL - 3508 TA Utrecht, the Netherlands
	 \and
                SRON National Institute for Space Research, Sorbonnelaan 2,
                NL - 3584 CA Utrecht, the Netherlands 
         \and
                Dept. of Astronomy, University of Maryland, College Park,
                MD 20742, U.S.A.
	 \and
                NASA Goddard Space Flight Center, Code 662, Greenbelt,
                MD 20771, U.S.A.
         \and
                Istituto di Astrofisica Spaziale e Fysica Cosmica 
                (CNR), Area Ricerca Roma Tor
                Vergata, Via del Fosso del Cavaliere, I - 00133 Roma, Italy
	}

\date{Received, accepted }

\abstract{
\bron\ is a persistent moderately bright X-ray source 1\fdg2 from the 
Galactic center. X-ray
observations with the Wide Field Cameras on BeppoSAX have for the
first time revealed the true nature of \bron: a 15-min long
thermonuclear flash was detected exposing the source as a neutron star
in a binary system. The flash was Eddington-limited constraining
the distance to between 5 and 8~kpc. We analyze BeppoSAX, ROSAT,
and RXTE data on \bron. The persistent 0.5--200~keV luminosity is close
to or less than 1\% of the Eddington limit which implies a rarely-seen
circumstance for thermonuclear flash activity.
 }

\maketitle 

\section{Introduction}
\label{intro}

\bron\ was discovered with the {\em Spacelab-2}/X-Ray Telescope
(SL2-XRT) which flew from 29 July to 6 August 1985 (Skinner et
al. 1987). SL2-XRT targeted the Galactic center a number of times for
a total of 7~hrs. \bron\ was detected with an average flux of
$7.7\times10^{-11}$~\ecs\ in 3 to 30 keV. No timing or
spectral features were noticed.  Skinner et al. also went back to data
from the {\em Einstein\/} galactic plane survey (Hertz \& Grindlay
1983) and found that \bron\ was clearly present. The {\em Einstein\/}
source was later given the name of 2E~1737.5-2817 and provided a
1\arcmin-accurate position.

Sakano et al. (2002) reported on all {\em ASCA}-GIS observations of a
$5\degr\times5\degr$ field centered on the galactic center. The
position of \bron\ (also designated AX~J1740.7-2818) was covered on 3
occasions in 1996, 1998, and 1999 with a total exposure time of
61.5~ksec and was detected every time with 0.7--10 keV fluxes
between 4.8 and 6.5$\times10^{-11}$~\ecs. The spectrum is consistent
with a constant absorbed power law with a photon index $\Gamma$ of
2.17$\pm0.07$ and $N_{\rm H}=(2.0\pm0.1)\times10^{22}$~cm$^{-2}$. The
flux varied up to 100\% on time scales of 10$^3$ to 10$^4$~s.

Because of being so close to the Galactic center, \bron\ was serendipitously
targeted in a number of X-ray wide-field monitoring programs such as
those carried out with Kvant/COMIS-TTM (In 't Zand 1992, Emelyanov et al.
2000) and Granat/ART-P (Pavlinsky et al. 1994). The source was never detected
in these programs. The upper limits are consistent with the ASCA and SL2-XRT
detections.

Despite that a neutron star hypothesis was suggested by the spectrum,
variability and flux, definite evidence was lacking; an
AGN hypothesis could not be ruled out (e.g., Sakano et al. 2002). This
has now changed. We present in this paper the first detection of an
X-ray burst from \bron\ which settles the nature of \bron. In addition
we discuss ROSAT and RXTE data.

\section{BeppoSAX-WFC observations}
\label{wfc}

\begin{figure}[t]
\psfig{figure=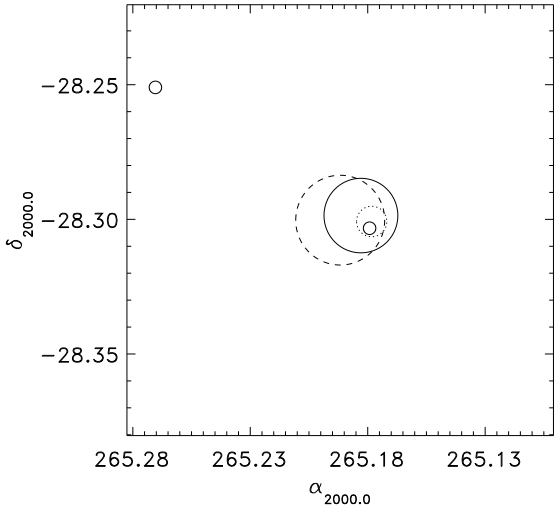,width=0.9\columnwidth,clip=t}
\caption{Locations of Einstein source (large dashed circle),
WFC-detected flare (large solid circle), RX~J1740.7-2818 (old position
/ dotted circle, and refined / small solid circle in middle),
2RXP~J174104.3-281504 (small solid circle in upper left corner). All
circles are at least 90\% confidence.
\label{figmap}}
\end{figure}

\begin{figure}[t]
\psfig{figure=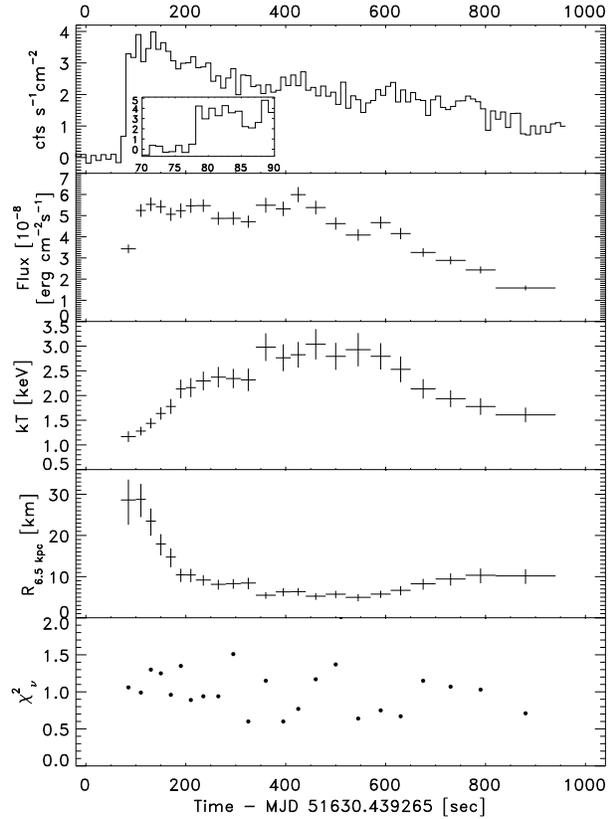,width=\columnwidth,clip=t}
\caption{Time history of WFC-measured full-bandpass photon flux for
the flare at 10~s resolution (top panel; the inset zooms in on the
flare onset at 1~s resolution), and of 4 parameters resulting from
modeling with black body radiation keeping $N_{\rm H}$ fixed at
1.9$\times10^{22}$~cm$^{-2}$ (see Sect.~\ref{nfi}): bolometric flux
(2nd panel), black body temperature (3rd panel), black body spherical
radius for an assumed distance of 6.5~kpc (4th panel), and reduced
$\chi^2$ value (5th panel).
\label{figwfclc}}
\end{figure}

The Wide Field Cameras (WFCs; Jager et al. 1997) on the BeppoSAX
satellite (Boella et al. 1997a) carry out a monitoring program in 2 to
28 keV of the galactic center since mid 1996. \bron\ is well inside
the $40\degr\times40\degr$ field of view and, through the 5\arcmin\
angular resolution, well-resolved from nearby sources. The monitoring
program till early 2002 totalled a net exposure of $\approx$70
days. \bron\ was never detected in individual observations above an
upper limit of roughly 2$\times10^{-10}$~\ecs\ (2-10 keV) except for a
single 15~min period on 27 March 2000 when a flare was detected from a
location consistent with the {\em Einstein\/} position of \bron\ (see
Fig.~\ref{figmap}).  Figure~\ref{figwfclc} (top panel) shows the time
history of the intensity. The event is characterized by a fast rise --
within one second -- and a factor-of-3.5 decay in the following
880~sec (or an e-folding decay time of 600~s). At that point a data
gap starts which lasts 1122~s after which \bron\ is not detected
anymore. Thus, the flare is shorter than 2000~s.

We modeled the spectrum of the flare successfully with pure black body
radiation, see Fig.~\ref{figwfclc}. This, and the softening of the
spectrum during the decay unambiguously identify this flare as a
so-called type-I X-ray burst originating from a thermonuclear flash on
the surface of a neutron star.  The bolometric fluence of the burst
before the data gap is $3.0\times10^{-5}$~erg~cm$^{-2}$.

The burst lasts fairly long, with a 2-min long radius expansion
phase at the start and a 6.5 min period after that when the flux and
temperature retain their maximum values. 
The radius expansion during the burst suggests that the flux reaches
near-Eddington values.  This allows a rough estimate of the
distance. For an Eddington limit between 2 and
4$\times10^{38}$~\lum\ (Kuulkers et al., in prep.), the peak flux
of $(6.0\pm0.5)\times10^{-8}$~\ecs\ yields a distance between 5
and 8~kpc.

\section{BeppoSAX-NFI observations}
\label{nfi}

A follow-up observation of the burst was carried out with the BeppoSAX
Narrow Field Instruments (NFI) half a year later, from 14 October 2000
01:44 UT to 15 October 2000 10:25 UT. Three NFI were turned on: the
low-energy concentrator spectrometer LECS (0.1--4 keV; Parmar et
al. 1997), the medium-energy concentrator spectrometer MECS (1.8--10.5
keV; Boella et al. 1997b), and the Phoswich Detector System PDS
(Frontera et al. 1997). The PDS is a collimating instrument with a
hexagonal field of view of diameter 1\degr3 to 1\fdg5; the LECS and
MECS are imaging devices with field of views of 40\arcmin\ and
30\arcmin\ diameter respectively and few arcmin resolution. The
exposure times were 14.0 ksec (LECS), 48.0 ksec (MECS) and 22.2 ksec
(PDS).

\begin{figure}[t]
\psfig{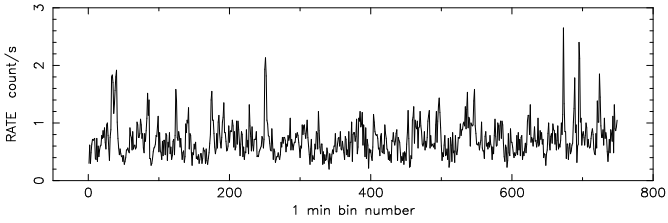}
\caption{Time history of photon rate in the MECS with a
resolution of 1~min. Twenty one data gaps have been cut out. No
background (estimated at 0.02 s$^{-1}$) was subtracted. The typical
1$\sigma$ error per data point is 0.1~c~s$^{-1}$.
\label{figmecslc}}
\end{figure}

\bron\ was clearly detected in the LECS and MECS
data. Figure~\ref{figmecslc} shows the light curve as derived from the
MECS data. The source exhibits variability with a maximum to minimum ratio
of 10 to 1 on typical time scales of a few minutes. The power
spectrum shows red noise (i.e., a power-law function with an index of
$-1.17\pm0.04$) that becomes indistinguishable from Poisson noise
beyond 1~Hz. There are no narrow features. The fractional rms is
$40\pm1$\% between 0.001 and 1~Hz.

A spectrum was generated by combining all LECS, MECS and PDS data. For
the data extraction, standard recipes were employed. The off-source
pointings of the rocking collimators of the PDS, employed to determine
background levels, were verified in WFC data to have no (bright)
sources in their field of views. The resulting spectrum could not be
modeled in an acceptable way, the reason being that too much flux
appears to be contained in the PDS data: all models mandate an
unacceptably strong step of a factor of $2.5\pm0.1$ between the MECS
and PDS data. We suspect that either the PDS picks up stray light from
the bright source GX~3+1 that is located 2\fdg36 from \bron, or that
other faint sources are present inside the field of view of the PDS
that confuse the signal from \bron\ although there are no other
sources detected in the (smaller) field of views of LECS and MECS. We
decided to exclude the PDS data. The LECS and MECS data can be
satisfactorily fit with an absorbed power law ($\chi^2_\nu=1.4$ for
$\nu=77$ and a systematic error of 3\%) with a photon index of
$2.12\pm0.04$ and $N_{\rm
H}=(1.9\pm0.1)\times10^{22}$~cm$^{-2}$. These numbers are consistent
with the ASCA ones. The observed flux is $6.3\times10^{-11}$~\ecs\ in
0.5--10 keV ($1.4\times10^{-10}$~\ecs\ unabsorbed). The PDS data show
that the 10-200 keV flux must be below $9.0\times10^{-11}$~\ecs, so
that the unabsorbed 0.5-200 flux must be $(1.4-2.3)\times10^{-10}$~\ecs.

\section{ROSAT-PSPC observations}
\label{rosat}

ROSAT carried out a number of pointings on 29 February and 1 March
1992 which cover \bron. All of these were with the Position-Sensitive
Proportional Counter (PSPC) in 0.5 to 2.4 keV.  We analyzed these data
because they represent the X-ray data with the highest angular
resolution and enable a check on source confusion. In the two
observations where \bron\ is almost on axis (obs ids rp400199 and
rp400200), we identify two sources close to \bron: 1RXS~J174043.1-281806
(=RX~J1740.7-2818) and 2RXP~J174104.3-281504. Only the former is
consistent with the {\em Einstein\/} position of \bron, see
Fig.~\ref{figmap}, and we identify it with \bron. The other is close
to Tycho-2 star 6839-218-1, whose colors are compatible with a G star
at about 150 pc.  The detected X-ray flux is compatible with such a
counterpart.  Using this object to determine the bore sight, and
combining both ROSAT observations, we obtain a position for
1RXS~J174043.1-281806 of $\alpha_{2000.0}=17^{\rm h}40^{\rm
m}43\fs00$, $\delta_{2000.0}=-28^{\rm o}18\arcmin11\farcs9$ or $l^{\rm
II}=359\fdg9731$, $b^{\rm II}=+1\fdg2481$. The positional accuracy is
limited by the statistical error of 2RXP~J174104.3-281504 to 8\farcs3
at 90\% confidence. This is a considerable improvement over the
previously determined radius of 20\farcs3. Motch et al. (1998)
attempted to find the optical counterpart of 1RXS~J174043.1-281806 in
the $I$-band image of this heavily reddened field but were not
successful. Within our refined position with 8\farcs3 radius, there
are still 11 stars in this image without an obvious candidate (as
identified, for instance, by Balmer line emission). We note that the
2MASS catalog shows a bright object within the error circle,
but it is too bright ($K$-dereddened 5.65) to be consistent with
\bron.  The 0.5--2.4 keV spectrum is consistent with the ASCA result.
The best-fit $N_{\rm H}$ for an absorbed power-law with $\Gamma=2.17$
is $(2.2\pm0.1)\times10^{22}$~cm$^{-2}$ with a flux of
7.2$\times10^{-12}$~\ecs.

\section{RXTE-PCA observations}
\label{pca}

\begin{figure}[t]
\psfig{figure=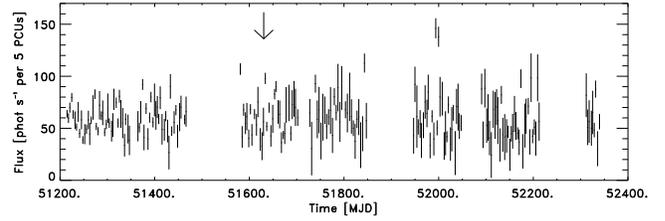,width=\columnwidth,clip=t}
\caption{Time history of the 2-13 keV photon rate of \bron\ during
206 RXTE-PCA bulge scans, from Feb 1999 to Mar 2002. One Crab unit is 10.4
kcts~s$^{-1}$ counted over all 5 proportional counter units (PCUs) that
make up the PCA. The arrow indicates the time of the WFC-detected burst.
\label{figbulge}}
\end{figure}

Since February 1999, the Proportional Counter Array (PCA) on RXTE has
performed scans to monitor a $16\degr\times16\degr$ region around the
Galactic center.  Scans occur twice weekly, for a total exposure of
approximately 20 seconds per source per observation, and allows
detection of source flux variations at the $\sim$1 mCrab level (Swank
\& Markwardt 2001).  The temporal coverage is uniform, except for
periods from November--January and in June, when the positions of the
Sun and anti-Sun cross the galactic center region.  Light curves are
derived for each known point source in the region.
Figure~\ref{figbulge} shows that for \bron. It
shows a fairly steady source with at most a factor of four
variability. The mean flux of $(60.3\pm0.6)$~c~s$^{-1}$ per 5~PCUs
translates, for a Crab spectrum like applies to \bron, to
$1.2\times10^{-10}$~\ecs\ in 2--10 keV (3.0$\times10^{-10}$~\ecs\
unabsorbed in 0.5--10 keV).  However, we note that there is
considerable diffuse galactic emission near the galactic center, which
can bias the point source light curves rates upwards by several tens
of c s$^{-1}$ per 5 PCUs.  Still, the data show no strong brightening
of \bron\ in 206 measurements.

\section{Discussion}
\label{discussion}

\begin{table}
\caption[]{History of X-ray flux from \bron\label{table1}}
\begin{tabular}{llllll}
\hline\hline
Data set   & Time   & Band$^\ast$ & Flux$^\dag$ & Flux$^\ddag$ in \\
           &        & (keV)    &           & 0.5--10 keV\\
\hline
SL2-XRT    & Jul 85 & 3--30    & 7.7  & $13\pm3$ \\
ROSAT-PSPC & Feb 92 & 0.5--2.4 & 0.72 & $10\pm1$ \\
ASCA-GIS   & Sep 96 & 0.7--10  & 5.7  & $13\pm3$ \\
ASCA-GIS   & Sep 98 & 0.7--10  & 6.5  & $14\pm3$ \\
ASCA-GIS   & Mar 99 & 0.7--10  & 4.8  & $11\pm2$ \\
BSAX-NFI   & Oct 00 & 0.5--10  & 6.3  & $14\pm3$ \\
\hline\hline
\end{tabular}
$^\ast$Bandpass in which flux originally was measured; $^\dag$Original
flux measurement in $10^{-11}$~\ecs; $^\ddag$Flux in 0.5--10 keV band,
as extrapolated by assuming power law with photon index $\Gamma=2.15$
and $N_{\rm H}=1.9\times10^{22}$~cm$^{-2}$. The errors are propagated
errors in $\Gamma$ of 0.1 and in $N_{\rm H}$ of
$0.1\times10^{22}$~cm$^{-2}$.
\end{table}

The detection of the type-I burst from \bron\ unambiguously identifies
the origin as a neutron star. By analogy with other type-I X-ray
bursters, it is almost certain that this neutron star is in a low-mass
X-ray binary (LMXB) with a sub-solar mass companion.  According to a
count based on the LMXB catalog by Liu et al. (2001) which was updated
with recent discoveries, \bron\ is the 69th X-ray burster (c.f., In 't
Zand 2001). Among a LMXB population of about 150, this means that at
least about half of all LMXBs contain neutron stars. Identifying the
companion star in the optical may prove difficult, because of its
proximity to the galactic center and the high reddening ($A_{\rm
V}\approx11$) implied by the value of $N_{\rm H}$.

Whenever the average 0.5--10 keV flux of \bron\ is measured with
sufficient accuracy (see Table~\ref{table1}) it is in a narrow range
of 1.0 to 1.4$\times10^{-10}$~\ecs. The NFI observation shows that
this may extrapolate to as much as $2.3\times10^{-10}$~\ecs\ in
0.5--200 keV. The RXTE data show that during several hundred brief
monitoring observations in 3 years the source never shows signs of
strong brightening. Given the distance, the 0.5--200 keV flux
translates to a luminosity of $5\times10^{35}$ to
$1.8\times10^{36}$~\lum, suggesting a bolometric luminosity that is
less than or similar to 1\% of the Eddington limit (as is also
suggested by a direct comparison between the peak flux of the
Eddington-limited burst and the persistent flux).

If one ignores `superbursts' (e.g., Kuulkers et al. 2002a), the burst
duration and fluence are among the largest (c.f., Tawara et al. 1984,
Lewin et al. 1984, Van Paradijs et al. 1990, Swank et al. 1977,
Kaptein et al. 2000, Kuulkers et al. 2002b). The amount of radiated
energy, if isotropic, is between 1.0 and $2.2\times10^{41}$~erg. If
the energy released per gram of accreted material is between 1.6 and
6$\times10^{18}$~erg, this is equivalent to $2\times10^{22}$ to
$1.4\times10^{23}$~g of accreted material. Given the persistent
luminosity of perhaps a few times 10$^{36}$~\lum, the mass accretion
rate may be 5 to 10$\times10^{15}$~g~s$^{-1}$ (or 1 to
2$\times10^{-10}$~M$_\odot$~yr$^{-1}$). This implies a waiting time
between bursts between 20 and 320 days, consistent with the detection
of just a single burst.  The long duration combined with the low
persistent flux suggests that the burst is caused by a thermonuclear
flash in a hydrogen-rich environment in a mass accretion regime of
$\dot{M}<2\times10^{-10}$~M$_\odot$yr$^{-1}$ (Fujimoto et al. 1981).

There is a slight similarity to sources like 2S~0918-549 (e.g., Jonker
et al. 2001) and 4U~1850-087 (e.g., Kitamoto et al. 1992). These are
also rare bursters and moderately bright sources that, foremost, show
broad emission features at about 0.7 keV attributable to an
overabundance of Ne which would be consistent with a hydrogen-depleted
donor in an ultracompact orbit (Juett et al.  2001). Our data are not
constraining with regards to the 0.7~keV feature due to the high value
of $N_{\rm H}$, but the hydrogen-rich environment inferred from the
burst duration suggests that the similarity is superficial. In this
respect we note that the burst in \bron\ is an order of magnitude
longer than in 2S~0918-549 and 4U~1850-087.

\acknowledgement JZ and EK acknowledge financial support from the
Netherlands Org. for Scientific Research (NWO).

\end{document}